\documentclass[12pt, epsf]{article}

\usepackage{epsfig}
\usepackage{amssymb}
\usepackage{graphicx}
\usepackage{color}
\usepackage{subfigure}

\begin{document}

\baselineskip 6mm
\renewcommand{\thefootnote}{\fnsymbol{footnote}}

\newcommand{\nc}{\newcommand}
\newcommand{\rnc}{\renewcommand}

%%%%%%%%%%%%%%%%%%%%%% Equation Numbering %%%%%%%%%%%%%%%%%%%%%%%
%\makeatletter \rnc{\theequation}{\thesection.\arabic{equation}}
%\@addtoreset{equation}{section} \makeatother

%%%%%%%%%%%%%%%%%%%%%%%%%%%%%%%%%%%%%%%%%%%%%%%%%%%%%%%%%%%%%%%%%
%                                                               %
%                NEW COMMANDS AND MACROS                        %
%                                                               %
%%%%%%%%%%%%%%%%%%%%%%%%%%%%%%%%%%%%%%%%%%%%%%%%%%%%%%%%%%%%%%%%%

\newcommand{\tcb}{\textcolor{blue}}
\newcommand{\tcr}{\textcolor{red}}
\newcommand{\tcg}{\textcolor{green}}

%%%%% Simplify some frequently used LaTeX commands %%%%%

\def\be{\begin{equation}}
\def\ee{\end{equation}}
\def\ba{\begin{array}}
\def\ea{\end{array}}
\def\bea{\begin{eqnarray}}
\def\eea{\end{eqnarray}}
\def\nn{\nonumber\\}

%%%%%  Temporary notation %%%%

\def\ct{\cite}
\def\la{\label}
\def\eq#1{(\ref{#1})}

%%% Greek letters %%%

\def\a{\alpha}
\def\b{\beta}
\def\g{\gamma}
\def\G{\Gamma}
\def\d{\delta}
\def\D{\Delta}
\def\ep{\epsilon}
\def\e{\eta}
\def\ph{\phi}
\def\Ph{\Phi}
\def\ps{\psi}
\def\Ps{\Psi}
\def\k{\kappa}
\def\l{\lambda}
\def\L{\Lambda}
\def\m{\mu}
\def\n{\nu}
\def\th{\theta}
\def\Th{\Theta}
\def\r{\rho}
\def\s{\sigma}
\def\S{\Sigma}
\def\ta{\tau}
\def\o{\omega}
\def\O{\Omega}
\def\pr{\prime}

%%%%% Mathematical Symbols

\def\half{\frac{1}{2}}

\def\goto{\rightarrow}

\def\na{\nabla}
\def\grad{\nabla}
\def\curl{\nabla\times}
\def\div{\nabla\cdot}
\def\pa{\partial}

\def\bra{\left\langle}
\def\ket{\right\rangle}
\def\lb{\left[}
\def\lc{\left\{}
\def\ls{\left(}
\def\lp{\left.}
\def\rp{\right.}
\def\rb{\right]}
\def\rc{\right\}}
\def\rs{\right)}
\def\cl{\mathcal{l}}

\def\vac#1{\mid #1 \rangle}

%%%%  Special symbol
\def\td#1{\tilde{#1}}
\def\check{ \maltese {\bf Check!}}

%%%%% Roman pont in math

\def\Tr{{\rm Tr}\,}
\def\det{{\rm det}}

%%%%% Special format

\def\bc#1{\nnindent {\bf $\bullet$ #1} \\ }
\def\ch {$<Check!>$ }
\def\ss {\vspace{1.5cm}}

\begin{titlepage}

%---------------- preprint number ---------------
\hfill\parbox{5cm} { }

\vspace{25mm}

\begin{center}
%------------------------ title ------------------------
{\Large \bf  Higher order WKB corrections to black hole entropy in brick wall formalism}

%---------------- authors and addresses ----------------
\vskip 1. cm
  {
  Wontae Kim$^{ab}$\footnote{e-mail : wtkim@sogang.ac.kr} and
  Shailesh Kulkarni$^a$\footnote{e-mail : skulkarnig@gmail.com}
  }

\vskip 0.5cm

{\it $^a\,$Center for Quantum Spacetime (CQUeST), Sogang University, Seoul 121-742, Korea}\\
{\it $^b\,$Department of Physics,, Sogang University, Seoul 121-742, Korea}\\

\end{center}

\thispagestyle{empty}

\vskip2cm

%----------------------- abstract ----------------------

\centerline{\bf ABSTRACT} \vskip 4mm

\vspace{1cm}

\noindent 
We calculate the statistical entropy of a quantum field with an arbitrary spin
propagating on the spherical symmetric black hole background  by using 
the brick wall formalism at higher orders in the WKB approximation. 
For  general spins, we find that the correction to the standard Bekenstein-Hawking 
entropy depends logarithmically on the area of the horizon. Furthermore, we apply this analysis  to the Schwarzschild and 
Schwarzschild-AdS black holes and discuss our results. 

\vspace{2cm}

%PACS numbers:

%\today

\end{titlepage}

\renewcommand{\thefootnote}{\arabic{footnote}}
\setcounter{footnote}{0}

%\tableofcontents
%%%%%%%%%%%%%%%%%%%%%%%%%%%%%%
%                                                                            %
%   Sec.  Introduction                                                       %
%                                                                            %
%%%%%%%%%%%%%%%%%%%%%%%%%%%%%%
\section{Introduction}
Thermodynamics and statistical mechanics of black holes are the most exciting and rapidly developing
areas of black hole physics. The analogy between thermodynamics and black holes was originally introduced
by Bekenstein  \cite{Bek1,Bek2} by assigning the area $A_{h}$ of the black hole to the entropy $S_{BH}$.
This correspondence was placed on the formal footing by  Hawking's discovery that black holes radiate thermally
 with the characteristic temperature ($T_{H}$) \cite{hawking1,hawking2},
\be
T_{H} = \frac{\hbar\kappa}{2\pi} \ls\frac{c}{k_{B}}\rs \la{eq:Hawktemp}
\ee 
where $\kappa$ is the surface gravity of the black hole and $\hbar, c, k_{B}$ denote the Planck constant,
the speed of light and the Boltzmann constant, respectively. Upon using the above expression for the temperature,
the Bekenstein-Hawking entropy is given by
\be
S_{BH} = \frac{k_{B}A_{h}}{4\ell^{2}_{p}}, \la{eq:BHentropy} 
\ee
with the Planck length, $\ell_{p}= \sqrt{G_{N}\hbar/c^3}$. 

  There are several approaches to derive the black hole entropy. In the Euclidean approach \cite{hawking3,york1,brown,
deAlwis:1994ej,deAlwis:1995cr},   using the analytic continuation to the Euclidean sector and imposing the Matsubara period 
$\beta = T_{H}^{-1}$,  the free energy and hence entropy have been obtained for the regular Euclidean solution of the vacuum 
Einstein equations.  In the context of the string theory, the statistical entropy  for extremal \cite{strominger1,khuri,strominger2,sen1,sen2}   
and near extremal \cite{callan,strominger3}  black holes is determined by explicit counting of the microstates  associated with  
the $D-branes$. Apart from these,  the calculation of the black hole entropy and the  corresponding quantum corrections have been studied 
by using various methods   like- spin networks \cite{ashtekar,partha}, entanglement between the degrees of freedom separated by Killing horizons  
\cite{srednicki,shanky,solodukhin4}, the conformal anomaly methods \cite{carlip1,carlip2,Cai:2009ua} and the quantum  tunneling approach 
{\cite{paddy3,bibhas1}}. Very recently,  the Euclidean approach  mentioned above, has been  implemented to derive the logarithmic corrections to 
Bekenstein-Hawking entropy  for extremal as well as non-extremal black holes in the various dimensions by incorporating the zero modes  \cite{sen3}.
 
   There is another efficient method, proposed by 't Hooft \cite{hooft1}, of computing the black hole entropy. 
 This formalism, commonly known as {\it brick wall model}, is the semi-classical approach wherein the gravitational
  field (metric) of the black hole is treated classically, while the remaining degrees of freedom, leaving
 outside the horizon, are handled quantum mechanically. The entropy of these quantum degrees of freedom, calculated via
  statistical mechanics, is identified with the entropy of the black hole. An important ingredient in the brick wall model
 is the boundary condition on the probe fields near the horizon. Any quantum fields near the black hole have a crucial
 property. Namely, the density of states of the quantum mechanical Hamiltonian of the fields blows out in the vicinity
 of the horizon. This results in the divergence of the statistical  entropy \cite{susskind,Mann:1990fk, meyers, emparan}.
  In the context of brick wall approach, this divergence can be controlled by putting a static spherically symmetric mirror near 
  the horizon, at which the fields are required to satisfy Dirichlet or Neumann boundary conditions \cite{hooft1}. 
  In other words,  the distance between the horizon and the mirror (brick wall) behaves as an ultra violate cut-off. 
   The canonical (statistical) entropy obtained by using this prescription is finite and agrees with Bekenstein-Hawking formula 
  (\ref{eq:BHentropy}).  This procedure also holds for any spacetime endowed with the horizon \cite{Padmanabhan:1986rs}.
 The status of the brick wall model have been elevated to formal level in \cite{israel}. 
  They found that  the expectation values of energy-momentum tensors for  quantum fields in the ground state  
 in the brick wall model matches exactly with the difference between Hartle-Hawking and Boulware states. 
 
   All the methods of computing the statistical entropy of the  black hole, in spite of their differences
in the underlying assumptions and methodology, correctly reproduce the same leading order result. However, the 
quantum corrections are generally different for different approaches. For instance, the coefficient 
of the logarithmic corrections to the entropy obtained from fluctuations around the  stable canonical ensemble \cite{parth2}
is different from  the one obtained using tunneling formalism \cite{bibhas1,sujoy1}. It is then natural to seek for
the extension of the brick wall model beyond the leading order. Such an extension of the brick wall approach 
up to sixth order in the WKB approximation has been reported in Ref. \cite{sriram1}. They showed that
for spherically symmetric black holes, in four dimensions as well as in six dimensions, the corrections to
 the brick wall entropy can be expressed as  $F(A_{h})\ln(A_{h}/\ell^{2}_{p})$ where $A_{h}$ is area of the horizon
 and the form of $F(A_{h})$ depends on the specific details of the black hole. 
  
  Although  the brick wall approach is implemented to obtain the entropy of a quantum scalar field in various 
background geometries and in diverse dimensions \cite{otherbackground}, 
relatively less attention  has been given for its generalization to other type of fields, {\it e.g}, fermions, photons or gravitons  
\cite{otherfields}. Any computation of the entropy for a quantum field with generic spin degrees of freedom,
 even at the leading order, seems to be difficult \cite{arbitspin}. It is therefore worthwhile to analyze the higher
 order WKB  corrections to the canonical entropy for arbitrary spins.  In this work, by using the formalism given in
 Ref.\cite{sriram1} we calculate, up to second order in the WKB approximation, the statistical entropy of a massless
 quantum field with the generic spin (scalar,vector,spinor etc.) propagating in the $3+1$ dimensional spherically
symmetric black hole background. For the general spins, we show that the functional dependence of entropy on the 
horizon area is the same for the leading as well as second order terms. However, the contribution to entropy coming from 
the second order term is appreciable compare to the leading order. The total entropy is obtained by adding the contributions 
from the leading and second order results.  

The paper is organized in the following manner. In the next section, we  briefly outline the method for extending the
brick wall entropy to higher order in the WKB approximation and generalize it for arbitrary spins.
 Section $3$ is devoted to explicit calculations of the leading order and second order brick wall entropies.
Applications of this approach to specific black holes are exhibited in section $4$.  We summarise our results in section $5$.     
\section{Brick Wall approach for arbitrary spin}
We consider a $3+1$ dimensional spherically symmetric black hole metric\footnote{Throughout the paper we shall work with
 the signature $(-,+,+,+)$, and $c=G=k_{B} =1$.}, 
\be
ds^{2} =  -g(r)dt^{2} + \frac{dr^{2}}{g(r)} + r^{2}(d\theta^{2} + \sin^{2}\theta d\phi^{2}) \la{eq:metric-1}
\ee
where the form of the metric coefficient $g(r)$ depends upon the specific black hole. The event horizon $r_{h}$ is defined by the condition
$g(r)=0$. For  a spherically symmetric black hole, the Hawking temperature is given by
\bea
 T_{H} = \frac{\hbar \kappa}{2\pi} = \frac{\hbar g'(r_{h})}{4\pi} \la{eq:kappaandTH}
\eea
where $\kappa$ is the surface gravity and  the prime denotes differentiation with respect to  the radial coordinate. 
We consider the massless minimally coupled field $\Phi_{p}$ with an arbitrary spin $|p|$ ($|p| =0,1/2,1\cdots$), propagating 
in the background gravitational field (\ref{eq:metric-1}). The equations of motion for $\Phi_{p}$ are given by \cite{Tueko,chandra} 
\be
{_{p}\Box} \Phi_{p} = 0 \la{eq:waveeq}
\ee
where ${_{p}\Box}$ is the generalized d'Alembertian operator
\bea 
{_{p}\Box} &=& -\frac{r^2}{g}\pa_{t}^{2} - 2 p r \ls 1-\frac{r \pa_{r}g}{2 g} \rs \pa_{t} + (r^2 g)^{-p} \ \pa_{r}\lb
 (r^2 g)^{p+1}\pa_{r}\rb  \nn
&& + \frac{1}{\sin\theta}\pa_{\theta}(\sin\theta \pa_{\theta}) + 2i p \frac{\cos\theta}{\sin^{2}\theta}\pa_{\phi}
 + \frac{1}{\sin^{2}\theta}\pa^{2}_{\phi} - (p^{2}\cot^{2}\theta
- p ) .
\eea 

\noindent For the static and spherically symmetric background, we can consider the ansatz
\be 
 \Phi_{p}(t,r,\theta,\phi) = e^{-i Et/\hbar}\bar{R}_{p}(r)S^{p}_{lm}(\theta,\phi).  \la{eq:Phiansatz}
\ee 
Substituting this in  Eq.(\ref{eq:waveeq}), we get  the radial equation
\be
 (r^2 g)^{-p} \ \frac{d}{dr}\lb (r^2 g)^{p+1}\frac{d\bar{R}_{p}(r)}{dr}\rb + \lb  \frac{r^2 \o^2}{g} - \l\rb \bar{R}_{p}(r)
 + 2i p r \lb 1- \frac{r}{2g}\frac{dg}{dr}\rb \bar{R}_{p}(r) = 0 , \la{eq:radialeq}
\ee 

\noindent while the spin weighted spherical harmonics $S^{p}_{lm}(\theta,\phi)$ satisfy \cite{frolov} 
\be
\lb \frac{1}{\sin\theta}\pa_{\theta}(\sin\theta \pa_{\theta}) + 2i p \frac{\cos\theta}{\sin^{2}\theta}\pa_{\phi}
 + \frac{1}{\sin^{2}\theta}\pa^{2}_{\phi} - (p^{2}\cot^{2}\theta - p) \rb S^{p}_{lm}(\theta,\phi) = - \l S^{p}_{lm}(\theta,\phi)
  \la{eq:angeq}
\ee
with
\be
\l = \ell(\ell +1) - p(p+1)    \  ; \     \ell \geq |p|  \la{eq:lambda}
\ee
and $E,l$ and $m$ are the energy, orbital angular momentum and azimuthal angular momentum associated 
with the given normal mode, respectively. We can recast (\ref{eq:radialeq}) into more appropriate form by transforming 
the dependent variable $\bar{R}_{p}(r)$ as
\be
\bar{R}_{p}(r) = R_{p}(r)\frac{1}{(r^2 g)^{\frac{p+1}{2}}}~  . \la{eq:R}  
\ee        
After performing some algebra, we get
\bea
 R''_{p}(r) & +&  \frac{1}{g^2}\lb \frac{E^2}{\hbar^2} - \frac{g \l}{r^2}\rb R_{p}(r)
 + \frac{2ip E}{\hbar} \lb \frac{1}{g}- \frac{rg'}{2 g^2}\rb R_{p}(r)\nn
&&   -(p+1)\lb \frac{g'}{r g}(p+1) + \frac{g'^{2}}{4 g^2}(p-1)+ \frac{g''}{2 g} + \frac{p}{r^2}\rb R_{p}(r) = 0 .  \la{eq:radialeq2}
\eea

  Our task is to solve Eq.(\ref{eq:radialeq2}) for each component\footnote{For a given value of spin $|p|$,  there are 
 $2p +1$ equivalent spin states. e.g for fermions we have $|p|= 1/2$ and $p= 1/2,-1/2$.)} of  the spin $|p|$ 
 and determine the number of radial modes of $\Gamma^{p}(E)$ of the field $\Phi_{p}$ with  the energy less
 than $E$. The total number of modes $\Gamma(E)$ is obtained by summing over all possible values of $p$.
When we perform the sum, the terms proportional to $ipE$ in Eq.(\ref{eq:radialeq2}) vanish. Therefore, 
 we can write  Eq.(\ref{eq:radialeq2}) as

\be
R''_{p}(r)+ \lb\frac{V^2_{1}(r)}{\hbar^2} - V_{2}(r)\rb R(r)= 0 \la{eq:radialeq3}
\ee
with the definitions:
\bea
V_{1}(r) &=& \frac{1}{g}\ls E^2 - \frac{g \lambda \hbar^2}{r^2} \rs^{1/2} \la{eq:V1}\nn
V_{2}(r) &=& (p+1)\lb \frac{g'}{r g}(p+1) + \frac{g'^{2}}{4 g^2}(p-1)+ \frac{g''}{2 g} + \frac{p}{r^2}\rb . \la{eq:V2}
\eea
 In almost all cases (except for $1+1$ dimensions) it is impossible to get an exact analytic solution of
Eq.(\ref{eq:radialeq3}) and the WKB approximation (at the leading order) is frequently used in the literature.
 We here compute the higher order WKB correction for a quantum field with a general 
spin $|p|$.  

  We start by considering the WKB ansatz for $R_{p}(r)$ in Eq.(\ref{eq:radialeq3}), 
\be
R_{p}(r) = \frac{1}{\sqrt{Q(r)}}e^{\frac{i}{\hbar} \int dr' Q(r')}  \la{eq:WKBansatz} 
\ee
where $Q(r)$ is an unknown phase. In order to analyze the higher order WKB approximation, we expand $Q(r)$ in powers of $\hbar^2$
\cite{sriram1,bender} as
\be
Q(r) = \sum_{i=0}^{\infty} Q_{2i}(r) \hbar^{2i} \la{eq:Qexpansion}. 
\ee 
After substituting Eqs.(\ref{eq:WKBansatz}) and (\ref{eq:Qexpansion}) into Eq.(\ref{eq:radialeq3}) and collecting the equal 
powers of $\hbar$, we get
\bea
Q_{0}(r) &=& \pm V_{1}(r) = \pm \frac{1}{g}\ls E^2 - \frac{g \lambda \hbar^2}{r^2} \rs^{1/2} , \la{eq:Q1}  \nn
Q_{2}(r) &=&  \frac{3Q_{0}^2(r)}{8Q^{3}_{0}(r)} - \frac{Q''_{0}(r)}{4Q^2_{0}(r)} - \frac{V_{2}(r)}{2Q_{0}(r)} , \la{eq:Q2}  \nn
Q_{4}(r) &=& - \frac{5Q^{2}_{2}(r)}{2Q_{0}(r)} - \frac{1}{4Q^{2}_{0}(r)}\ls 4Q_{2}(r)V_{2}(r) + Q''_{2}(r)\rs \nn &&
+ \ls\frac{3Q'_{2}(r)V'_{1}(r) - Q_{2}(r)V''_{1}(r)}{4V^{3}(r)}\rs  \la{eq:Q4}
\eea
for $\hbar^{0}, \hbar^{2}$ and $\hbar^4$ orders respectively. In fact, all the higher order functions can be expressed 
in terms of potential $V_{1}(r)$ and $V_{2}(r)$. Upon using the semi-classical quantization scheme \cite{hooft1} together
with the series expansion of $Q(r)$, we write the expression for the density of states $\Gamma(E)$ with energy less that $E$ 
as
\be
\Gamma(E) = \sum_{i=0}^{\infty}\Gamma_{2i}(E) . \la{eq:totdensitystates} 
\ee
Here $\Gamma_{2i}(E)$ denotes the number of states at $i^{th}$ order and it is given by

\be
\Gamma_{2i}(E) = \ls\frac{\hbar^{(2i-1)}}{\pi}\rs \sum_{p=1}^{2|p|+1} \int_{r_{h}+\ep}^{L} dr \int_{|p|}^{\ell_{m}} d\ell
 (2\ell + 1) Q_{2i}(r,\ell, p) .
\la{eq:DSi} 
\ee

We have regularized our computations by introducing the ultra violate and infra red cut-off \cite{hooft1}
such that the quantum field $\Phi_{p}$ satisfy:  $\Phi_{p}(r \leq r_{h}+\ep)=\Phi_{p}(r \geq L)=0$ with $\ep\ll 1$ 
and $L \gg r_{h}$. There are few points which we would like to emphasize at this stage. Normal modes of the quantum 
field in the region close to horizon undergoes an infinite blue shift and eventually gives divergent contribution to the 
density of states. In fact, this feature is always present when we study the quantum field theory in any spacetime with 
horizon \cite{paddy1,paddy2,spindel}. The ultra violate cut-off introduced above precisely regularize this divergence. 
On the other hand, for the massless fields the large distance behavior of the density of states is completely well defined. 
The free energy of the quantum field vanishes in $L\rightarrow \infty$ limit.  However, for massive case we get 
infra red (large distance) divergence. This divergence can be removed either by restricting the upper limit of radial 
integration or by introducing negative cosmological constant \cite{winstanley}. Another point which we would like to 
mention is that, in Eq.(\ref{eq:DSi}) we have replaced the sum over $\ell$ values by the
corresponding integral. The maximum of the $\ell$ integral is chosen such that $Q_{0}(r,\ell,p)$ is real. Such an approximation
 of replacing sum by integration is valid for the leading as well as higher order WKB analysis. For the general spin, situation
 is slightly different. In that case, there exist a lower limit on the $\ell$ integration and it is related to the consistency of the
 eigenspectra of the spin weighted spherical harmonics \cite{Tueko,breuer} (for more details, see \cite{morse} ).
 This modification complicates the higher order  of the WKB analysis.

  Using the density of states (\ref{eq:DSi}),  we can compute the free energy for bosons($-$)
 and fermions ($+$) from their standard expressions,
\bea
F^{\pm}_{2i} &=&  - \int_{0}^{\infty} dE \ \frac{\Gamma_{2i}(E)}{e^{\beta E} \pm 1} \la{eq:freeengbf}
\eea   
where $\beta$ is the inverse of the Hawking temperature. The statistical entropy associated with the free energy
 $F^{\pm}_{2i}$  is given by 
\be 
S^{\pm}_{2i} = \beta^{2} \ls\frac{\pa F_{2i}^{\pm}}{\pa \beta}\rs . \la{eq:entropybf}
\ee
\section{Leading and Second order computations}
In this section, we shall compute the statistical entropy of the quantum field with spin $|p|$ 
propagating on the spherically symmetric black hole background at the leading and second 
order in the WKB approximation. The total entropy is obtain by combining the
 leading and second order results for the corresponding species.  
\\
\noindent {\bf Leading order:}\\
From  Eq.(\ref{eq:DSi}), we have the following leading order expression, 
\be
\Gamma_{0}(E) = \frac{1}{\hbar\pi}\sum_{p} \int_{r_{h}+\ep}^{L} dr \int_{|p|}^{\ell_{m}} {d\ell} (2\ell + 1) 
\frac{1}{g(r)}\ls E^2 - \frac{g(r) \lambda \hbar^2}{r^2} \rs^{1/2}\la{eq:DS01}   
\ee 
where $\l$ is given by Eq.(\ref{eq:lambda}). It is convenient to work with variables $\tilde\l(r)$, $\tilde E$
and $G(\tilde{\l},\tilde{E})$, which are defined as
\bea
 \tilde\l &=& \frac{g \l\hbar^2}{r^2} = \frac{g\hbar^2}{r^2}(\ell(\ell+1)-p(p+1)), \la{eq:lambdatilde}\\
 \tilde E &=& E^{2}, \la{eq:Etilde}\\
 G(\tilde{\l},\tilde{E}) &=& \ls{\tilde{E} - \tilde{\l}}\rs^{1/2}. \la{eq:Gtilde}
\eea
Using these, we rewrite the density of states as
\be
\Gamma_{0}(E) = \frac{1}{\pi\hbar^3}\int_{r_{h}+\ep}^{L} dr \ls\frac{r^2}{g^{2}(r)}\rs \lb\sum_{p} 
\int_{\tilde\l_{0}}^{\tilde\l_{max}}
d\tilde\l  G(\tilde{\l},\tilde{E})\rb.
\ee 
The maximum of $\tilde{\l}$ is located where the function $G(\tilde{\l},\tilde{E})$ vanishes 
and the lower limit is dictated by the spin $|p|$. Using Eqs.(\ref{eq:lambda}) and (\ref{eq:Gtilde}), we find
\be
 \tilde\l_{max} = \tilde{E}  \ ; \  \l_{0} = \frac{g(r)\hbar^2}{r^2}\lb |p|(|p|+1) - p(p+1)\rb. \la{eq:lambdamaxlambdamin}
\ee
After performing the $\tilde{\l}$ integral and summing over $p$, we obtain
\be
\Gamma_{0}(E) = \frac{2}{3\pi\hbar^3}(2|p|+1) \tilde{E}^{3/2}\int^{L}_{r_{h}+\ep}  \frac{r^2}{g^{2}(r)} . \la{eq:DS02}
\ee
Substituting this in Eq.(\ref{eq:freeengbf}) and integrating over $E$, yields
\bea 
F_{0}^{-} &=& - \frac{2}{45}\ls\frac{\pi^3 (2|p|+1)}{\hbar^3 \beta^{4}}\rs \int^{L}_{r_{h}+\ep} \frac{r^2}{g^{2}(r)} , \la{eq:freeeng0b}\\
F_{0}^{+} &=& - \frac{14}{360}\ls\frac{\pi^3 (2|p|+1)}{\hbar^3 \beta^{4}}\rs \int^{L}_{r_{h}+\ep} \frac{r^2}{g^{2}(r)} \la{eq:freeeng0f}
\eea
for bosons and fermions, respectively. \\

Now,  the radial integrations are evaluated by expanding the metric near the horizon, 
\be
g(r) \approx g'(r_{h})(r-r_{h}) + \frac{g''(r_{h})}{2}(r-r_{h})^2 + \mathcal{O}((r-r_{h})^3). \la{eq:metricexpansion} 
\ee
At the quadratic order in the metric expansion, we get 
\bea
F^{-}_{0} &=&  - \frac{1}{45}\ls\frac{\pi^3 (2|p|+1)}{\hbar^3 \beta^{4}}\rs \lb\frac{r^{2}_{h}}{\tilde{\ep}^{2}\kappa^{3}}
+ \ls\frac{r^{2}_{h}g''(r_{h})}{4\kappa^3} - \frac{r_{h}}{\kappa^2}\rs \ln \ls\frac{r^{2}_{h}}{\tilde{\ep}^{2}}\rs\rb , \la{eq:freeeng01b}\\
F^{-}_{0} &=&  - \frac{7}{8}\frac{1}{45}\ls\frac{\pi^3 (2|p|+1)}{\hbar^3 \beta^{4}}\rs \lb\frac{r^{2}_{h}}{\tilde{\ep}^{2}\kappa^{3}}
+ \ls\frac{r^{2}_{h}g''(r_{h})}{4\kappa^3} - \frac{r_{h}}{\kappa^2}\rs \ln \ls\frac{r^{2}_{h}}{\tilde{\ep}^{2}}\rs\rb \la{eq:freeeng01f}
\eea
where we have used the coordinate invariant cut-off (proper distance of the brick wall from the horizon) 
\be 
\tilde{\ep}= \sqrt{\frac{4\ep}{g'(r_{h})}} \la{eq:handhc} ,
\ee
and $\kappa$ is the surface gravity given by Eq.(\ref{eq:kappaandTH}). The canonical entropy computed from Eq.(\ref{eq:entropybf}) 
becomes 
\bea
S^{-}_{0} &=& f^{-}(p) \frac{r^2_{h}}{90\tilde{\ep}^{2}} -  f^{-}(p)\lb \frac{r^{2}_{h}g''(r_{h})}{360}
-\frac{r_{h}\kappa}{90}\rb\ln\ls\frac{r^{2}_{h}}{\tilde{\ep}^{2}}\rs , \la{eq:entropy0b} \\
S^{+}_{0} &=& f^{+}(p) \frac{7}{8}\frac{r^2_{h}}{90\tilde{\ep}^{2}} - f^{+}(p)\frac{7}{8}\lb \frac{r^{2}_{h}g''(r_{h})}
{360}-\frac{r_{h}\kappa}{90}\rb\ln\ls\frac{r^{2}_{h}}{\tilde{\ep}^{2}}\rs .  \la{eq:entropy0f}
\eea
Here $f^{\pm}(p)=(2|p|+1)$ is the spin degeneracy factor for fermions and bosons respectively.
 Our leading order results are in agreement with that of given in the literatures  \cite{otherfields,arbitspin}. 
\\

\noindent {\bf Second order analysis:} \\
We now calculate the density of states of the quantum field with energy less than $E$ at the second order ($i=1$) in 
the WKB approximation. Upon using Eqs.(\ref{eq:lambdatilde}),(\ref{eq:Etilde}) and (\ref{eq:Gtilde})  we can write Eq.(\ref{eq:DSi})  as
\be
\Gamma_{2}(\tilde{E}) = \ls\frac{1}{\pi\hbar}\rs \int_{r_{h}+\ep}^{L} dr \ls\frac{r^2}{g(r)}\rs \lb I(\tilde{E},r) + J(\tilde{E},r)\rb
 \la{eq:DS2}  
\ee
where 
\bea
I(\tilde{E},r) &=& \sum_{p=1}^{2|p|+1} \int_{\tilde{\l}_{0}}^{\tilde{\l}_{m}} d\tilde{\l} \ls\frac{Q^{(0)}_{2}(r)}
{G(\tilde{\l},\tilde{E})} + \frac{\tilde{\l}}{G^{3}(\tilde{\l},\tilde{E})} Q_{2}^{(1)}(r) + \frac{\tilde{\l}^{2}}
{G^{5}(\tilde{\l},\tilde{E})} Q_{2}^{(2)}\rs , \la{eq:I}\\
 J(\tilde{E},r)&=& \sum_{p=1}^{2|p|+1} \int_{\tilde{\l}_{0}}^{\tilde{\l}_{m}} d\tilde{\l} \frac{\bar{Q}^{(0)}(r)}
 {G(\tilde{\l},\tilde{E})},
 \la{eq:J}
\eea
and we have defined 
\bea
Q_{2}^{(0)}(r) &=& -\frac{g'(r)}{2r}, \la{eq:Q20}\\
Q_{2}^{(1)}(r) &=& \frac{g'^{2}(r)}{8g(r)} - \frac{3g'(r)}{4r} + \frac{3g(r)}{4r^{2}}+\frac{g''(r)}{8}, \la{eq:Q21}\\
Q_{2}^{(2)}(r)&=& \frac{5}{32}\frac{g'^{2}(r)}{g(r)} - \frac{5}{8}\frac{g'(r)}{r} + \frac{5}{8}\frac{g(r)}{r^2} \la{eq:Q22},\\
\bar{Q}_{2}^{(0)}(r) &=& -p\lb(p+2)\frac{g'(r)}{2r} + p\frac{g'^{2}(r)}{8g(r)}+ (p+1)\frac{g(r)}{2r^{2}}+\frac{g''(r)}{4}\rb .
 \la{eq:Q20bar}
\eea 
It is worthwhile to point out that  Eqs.(\ref{eq:Q20}),(\ref{eq:Q21}) and (\ref{eq:Q22}) have the similar structures to those 
of given in Ref.\cite{sriram1}. While  the additional piece $\bar{Q}^{(0)}_{2}(r)$ depends explicitly on the spin orientation.
This term would vanish for $p=0$ (scalar). Therefore, we expect that the $\tilde{\l}$ integration over the terms 
$Q_{2}^{(0)},Q_{2}^{(1)}$ and $Q_{2}^{(2)}$ should yield the similar forms  given in Ref. \cite{sriram1}.  
However, the corresponding  integral over $\bar{Q}_{2}^{(0)}$ gives completely different structure. 
We shall evaluate these integrals one by one. 
   
  The definitions (\ref{eq:Etilde}) and (\ref{eq:Gtilde}) allow us to write inverse powers of
$G(\tilde{\l},\tilde{E})$ appearing in Eq.(\ref{eq:DS2}) as the partial differentials with respect to $\tilde{E}$. 
By successive applications of  the Leibniz rule of integral calculus and by interchanging the order of $\tilde{\l}$ integration with $\tilde{E}$
 differentiation, we obtain
\bea
I(\tilde{E},r) &=& (2|p|+1) \tilde{E}^{1/2}\lb 2Q^{(0)}_{2}(r)- 4Q^{(1)}_{2}(r) + \frac{16}{3} Q^{(2)}_{2}(r)\rb \nn
&& + \lb \frac{4\tilde{E}}{G(\tilde{\l},\tilde{E})}-\frac{8}{3}\ls\frac{\pa}{\pa \tilde{E}}\ls\frac{\tilde{E}^2}
{2G(\tilde{\l},\tilde{E})}\rs - \frac{\tilde{E}^{2}}{4G(\tilde{\l},\tilde{E})}\rs
\rb_{\tilde{\l}=\tilde{E}}. \la{eq:I1}
\eea
In getting the above expression, we have used 
\bea
\sum_{p=1}^{2|p|+1} (\tilde{E}-\tilde{\l}_{0})^{3/2}(2\tilde{E} + 3\tilde{\l}_{0}) &=& 2\tilde{E}^{5/2}(2|p|+1) ,\nn
\sum_{p=1}^{2|p|+1} (\tilde{E}-\tilde{\l}_{0})^{3/2} (8\tilde{E}^2 +12\tilde{E}\tilde{\l}_{0} + 15\tilde{\l}_{0}^2) &=& 
8 \tilde{E}^{7/2}(2|p|+1)\nonumber
\eea
where $\tilde{\l}_{0}$ is given by Eq.(\ref{eq:lambdamaxlambdamin}).\\

\noindent Next, we substitute $\bar{Q}_{2}^{(0)}$ in Eq.(\ref{eq:J}) and evaluate the integral to get
\bea
J(\tilde{E},r) = -2\sum_{p=1}^{2|p|+1}(\tilde{E}-\tilde{\l}_{0})^{1/2} p\lb(p+2)\frac{g'(r)}{2r} +
p\frac{g'^{2}(r)}{8g(r)}+ (p+1)\frac{g(r)}{2r^{2}}+\frac{g''(r)}{4}\rb . 
\eea
As mentioned before, this term vanishes for $|p|=0$. For $|p|>0$, we can execute the sum
by noting that
\bea
\sum_{p=1}^{2|p|+1} p (\tilde{E}-\tilde{\l}_{0})^{1/2} &=& \tilde{E}^{1/2} (1+|p|)(1+2|p|) , \nn
\sum_{p=1}^{2|p|+1} p^{2}(\tilde{E}-\tilde{\l}_{0})^{1/2}  &=& \frac{1}{3}\tilde{E}^{1/2} (1+|p|)(1+2|p|)(3+4|p|).\nonumber
\eea
Using the above formulae and introducing the theta function,  we write 
\bea
 J(\tilde{E},r) &=& -\tilde{E}^{1/2} \tilde{J}(r)\Theta(|p|)  \la{eq:J2}
\eea 
where
\be
 \tilde{J}(r)=(1+|p|)(1+2|p|)\lb (9+4|p|)\frac{g'(r)}{3r} + (3+4|p|)\frac{g'^{2}(r)}{12g(r)} 
 + (3+2|p|)\frac{2g(r)}{3r^{2}}+\frac{g''(r)}{2}\rb . \la{eq:Jtilde}
\ee
Note that $\Theta(|p|)$ for $|p|=0$ (scalar) vanishes, while  it is unity for $|p| \ne 0$.\\

  We should emphasize that  when $\tilde{\l}= \tilde{\l}_{max}=\tilde{E}$ (turning point) the function $G(\tilde{E},\tilde{\l})$ 
vanishes and all the derivatives with respect to $\tilde{E}$ diverge. Consequently, integration of the terms like $G^{-3}\tilde{\l}$ and 
$G^{-5}\tilde{\l}^{2}$ in Eq.(\ref{eq:I}) leads to the finite as well as diverging contributions. The divergent part is given by the last term
in Eq.(\ref{eq:I1}). This divergence is due to the fact that the WKB approximation is not valid as we move closer to the turning point.
On the other hand, we observe that the result for $J(\tilde{E},r)$ contains only finite term.
Thus, the structure of the divergent terms appearing for scalar particle \cite{sriram1} remain unchanged even for the generic spin.  

  Then, the  density of states $\Gamma_{2}(E)$ can be obtained by collecting the finite part of Eqs.(\ref{eq:I1}) and (\ref{eq:J2})
and substituting them into Eq.(\ref{eq:DS2}), 
\bea
\Gamma_{2}(E) &=& \frac{E(2|p|+1)}{\pi\hbar}\int_{r_{h}+\ep}^{L}dr \frac{r^2}{g(r)}\lc \ls 2Q^{(0)}_{2}(r)- 4Q^{(1)}_{2}(r)+ \frac{16}{3}
 Q^{(2)}_{2}(r)\rs \right.\la{eq:DS22}\\ 
&& \left. \frac{}{} - \Theta(|p|)(1+|p|)\ls \frac{(9+4|p|)g'(r)}{3r} + \frac{(3+4|p|)g'^{2}(r)}{12g(r)} + \frac{2(3+2|p|)g(r)}{3r^{2}}
+\frac{g''(r)}{2}\rs\rc . \nonumber 
\eea
Substituting the above expression in Eq.(\ref{eq:freeengbf}) and integrating over $E$, yields  
\bea
F^{-}_{2} &=& -\frac{(2|p|+1)\pi}{6\beta^2 \hbar}\int_{r_{h}+\ep}^{L}dr \frac{r^2}{g(r)}\lc \ls 2Q^{(0)}_{2}(r)-
 4Q^{(1)}_{2}(r)  + \frac{16}{3} Q^{(2)}_{2}(r)\rs \right.\\ 
&& \left. \frac{}{} - \Theta(|p|)(1+|p|)\ls \frac{(9+4|p|)g'(r)}{3r} + \frac{(3+4|p|)g'^{2}(r)}{12g(r)} + \frac{2(3+2|p|)g(r)}{3r^{2}}
+\frac{g''(r)}{2}\rs\rc , \nonumber 
\eea
\bea
F^{+}_{2} &=& -\frac{(2|p|+1)\pi}{12\beta^2 \hbar}\int_{r_{h}+\ep}^{L}dr \frac{r^2}{g(r)}\lc \ls 2Q^{(0)}_{2}(r)-
 4Q^{(1)}_{2}(r)  + \frac{16}{3} Q^{(2)}_{2}(r)\rs \right.\\ 
&& \left. \frac{}{} - \Theta(|p|)(1+|p|)\ls \frac{(9+4|p|)g'(r)}{3r} + \frac{(3+4|p|)g'^{2}(r)}{12g(r)} + \frac{2(3+2|p|)g(r)}{3r^{2}}
+\frac{g''(r)}{2}\rs\rc . \nonumber 
\eea        

\noindent Finally, by using the metric expansion (\ref{eq:metricexpansion}) and performing the radial integration, we get
\bea
F^{-}_{2} &=& - \frac{(2|p|+1)\pi}{6\beta^2 \hbar} \lc \lb\frac{4r^{2}_{h}}{3\tilde{\ep}^{2}g'(r_{h})} - \ls\frac{2r_{h}}{3}+
\frac{g''(r_{h})r^{2}_{h}}{6g'(r_{h})}\rs \ln\ls\frac{r^{2}_{h}}{\tilde{\ep}^{2}}\rs\rb \right. \la{eq:freeengb-f}
\\
&& \left. +\Theta(|p|)\ls 1+|p|\rs \frac{}{} \lb \frac{(3+4|p|)}{6}\ls\frac{r^2_{h}}{\tilde{\ep}^{2}g'(r_{h})}\rs -
\ls\frac{r_{h}}{4}(7+4|p|) + \frac{r^{2}_{h}g''(r_{h})}{24g'(r_{h})}(9+4|p|)\rs \right.\right. \nn
&& \left. \left. \frac{}{} \times\ln\ls\frac{r^{2}_{h}}{\tilde{\ep}^{2}}\rs\rb\rc , \nonumber\\
F^{+}_{2} &=& - \frac{(2|p|+1)\pi}{12\beta^2 \hbar} \lc \lb\frac{4r^{2}_{h}}{3\tilde{\ep}^{2}g'(r_{h})} - \ls\frac{2r_{h}}{3}+
\frac{g''(r_{h})r^{2}_{h}}{6g'(r_{h})}\rs \ln\ls\frac{r^{2}_{h}}{\tilde{\ep}^{2}}\rs\rb \right.  \la{eq:freeengf-f}
\\
&& \left. +\Theta(|p|)\ls1+|p|\rs \frac{}{} \lb \frac{(3+4|p|)}{6}\ls\frac{r^2_{h}}{\tilde{\ep}^{2}g'(r_{h})}\rs -
 \ls\frac{r_{h}}{4}(7+4|p|) + \frac{r^{2}_{h}g''(r_{h})}{24g'(r_{h})}(9+4|p|)\rs \right. \right. \nn
&& \left.\left. \frac{}{} \times\ln\ls\frac{r^{2}_{h}}{\tilde{\ep}^{2}}\rs\rb\rc . \nonumber
\eea
\noindent Thus, the canonical entropy at the second order in the WKB approximation for the quantum field of spin $|p|$ 
is given by 
\bea
S^{-}_{2} &=&  f^{-}(p)\lc\lb\frac{r^{2}_{h}}{9\tilde{\ep}^{2} }- \ls\frac{\kappa r_{h}}{9}+
\frac{g''(r_{h})r^{2}_{h}}{72}\rs \ln\ls\frac{r^{2}_{h}}{\tilde{\ep}^{2}}\rs\rb \right. \la{eq:entropy2b}
\\
&& \left. +\Theta(|p|)\ls1+|p|\rs \frac{}{} \lb \frac{(3+4|p|)}{72}\ls\frac{r^2_{h}}{\tilde{\ep}^{2}_{c}}\rs -
 \ls\frac{r_{h}\kappa}{24}(7+4|p|) + \frac{r^{2}_{h}g''(r_{h})}{288}(9+4|p|)\rs \right.\right. \nn
&& \left.\left. \frac{}{} \times\ln\ls\frac{r^{2}_{h}}{\tilde{\ep}^{2}}\rs\rb\rc , \nonumber
\eea
%\newpage
\bea
S^{+}_{2} &=&  f^{+}(p)\lc\lb\frac{r^{2}_{h}}{18\tilde{\ep}^{2}} - \ls\frac{\kappa r_{h}}{18}+
\frac{g''(r_{h})r^{2}_{h}}{144}\rs \ln\ls\frac{r^{2}_{h}}{\tilde{\ep}^{2}}\rs\rb \right. \la{eq:entropy2f}
\\
&& \left. +\Theta(|p|)\ls1+|p|\rs \frac{}{} \lb \frac{(3+4|p|)}{144}\ls\frac{r^2_{h}}{\tilde{\ep}^{2}}\rs -
 \ls\frac{r_{h}\kappa}{48}(7+4|p|) + \frac{r^{2}_{h}g''(r_{h})}{576}(9+4|p|)\rs \right.\right. \nn
&& \left.\left. \frac{}{} \times\ln\ls\frac{r^{2}_{h}}{\tilde{\ep}^{2}}\rs\rb\rc . \nonumber
\eea 
 Eqs.(\ref{eq:entropy2b}) and (\ref{eq:entropy2f}) constitute the main result.  For the  scalar field, our result is consistent with 
that given in Ref. \cite{sriram1}.  The canonical entropies (\ref{eq:entropy2b}) and (\ref{eq:entropy2f}) contain the quadratic as well as 
logarithmic divergent  terms in the ultra violate regime of $\tilde{\ep}\rightarrow 0$.  However, these divergences are even present
 at the leading order of the WKB approximations (\ref{eq:entropy0b}) and (\ref{eq:entropy0f}). Various approaches have been suggested 
 in the literature for regularizing these divergences \cite{susskind,meyers,emparan}, and it was shown that  they can be   absorbed into 
the renormalization of the coupling constants appearing in the one-loop effective action \cite{solodukhin1,solodukhin2,solodukhin3,
  fursaev1,shimomura}.
 
  We now write the expression for the total canonical entropy  up to the second order in the WKB approximation  by adding the 
contributions from the leading (\ref{eq:entropy0b},\ref{eq:entropy0f}) and second (\ref{eq:entropy2b},\ref{eq:entropy2f}) 
  order expressions,
 \be
 S^{\pm} = S^{\pm}_{0} + S^{\pm}_{2}. \la{eq:totentropyfb} 
 \ee 
 It is worthwhile to point out that  the invariant cut-off $\tilde{\ep}$ can be adjusted to match the leading 
 order term with the standard Bekenstein-Hawking entropy (\ref{eq:BHentropy}). For the scalar field, the degeneracy 
 factor $f^{-}(p=0)$ is unity. Thus, by setting $\tilde{\ep}^{2}_{sc} =(11\ell^{2}_{p}/90\pi)$ the standard expression 
 for $S_{BH}$ , at leading order, is obtained \cite{sriram1}. In general, the value of the invariant cut-off depends 
 upon the type of the field. The massless spin-$1/2$ (Weyl fermions) has only one polarization direction, and in this 
 case we set $\tilde{\ep}^{2}_{wf}= (169\ell^{2}_{p}/1440\pi)$.  While  the appropriate invariant cut-off length 
 for electromagnetic field (photon) is $\tilde{\ep}^{2}_{ph} =(57\ell^{2}_{p}/90\pi)$. Thus, we write the corresponding expressions for 
the  canonical entropy, 
  \bea
 S^{-}_{sc} &=& \frac{A_{h}}{4\ell^{2}_{p}}  -\lb \frac{\kappa}{10 r_{h}}+\frac{g''(r_{h})}{{60}}
  \rb \frac{A_{h}}{4\pi} \ln\ls\frac{A_{h}}{\ell^{2}_{p}}\rs , \la{eq:entropyupto2-scalar}\\
 S^{+}_{wf} &=& \frac{A_{h}}{4\ell^{2}_{p}}  - \lb \frac{471}{360} \ls\frac{\kappa }{r_{h}}\rs - \frac{657}{4320}g''(r_{h})\rb
   \frac{A_{h}}{16\pi} \ln\ls\frac{A_{h}}{\ell^{2}_{p}}\rs , \la{eq:entropyupto2-wf}\\
 S^{-}_{ph} &=& \frac{A_{h}}{4\ell^{2}_{p}}  - \lb \frac{122 }{15}\ls\frac{\kappa }{r_{h}}\rs + \frac{77}{90} g''(r_{h})\rb
 \frac{A_{h}}{16\pi} \ln\ls\frac{A_{h}}{\ell^{2}_{p}}\rs  \la{eq:entropyupto2-mx}  
 \eea 
for the massless scalar,  the fermion and the electromagnetic field, respectively.     
\section{Specific examples}
 We apply the previous analysis for the Schwarzschild and  Schwarzschild-AdS black holes and write
corresponding expressions for the brick wall entropy up to the second order in the  WKB approximation. \\

\noindent {\bf Schwarzschild black hole:}\\
The metric coefficient $g(r)$ and the surface gravity $\kappa$ for the Schwarzschild black hole of mass $M$ are  given by
\bea
g(r) &=&  1- \frac{r_{h}}{r}  , \\
\kappa &=& \frac{1}{2r_{h}} \la{eq:surfgrav-schwarz}
\eea
where $r_{h}= 2M$. 
Substituting these into Eqs.(\ref{eq:entropyupto2-scalar})-(\ref{eq:entropyupto2-mx}), we get  
\bea
 S^{-}_{sc} =  \frac{A_{h}}{4\ell^{2}_{p}} - \frac{1}{60}\ln\ls\frac{A_{h}}{\ell^{2}_{p}}\rs  , \quad   \  S^{+}_{wf} =
 \frac{A_{h}}{4\ell^{2}_{p}} - \frac{23}{96}\ln\ls\frac{A_{h}}{\ell^{2}_{p}}\rs , \quad   \ 
 S^{-}_{ph} =  \frac{A_{h}}{4\ell^{2}_{p}} - \frac{131}{144}\ln\ls\frac{A_{h}}{\ell^{2}_{p}}\rs . \la{eq:entropymxschwr} 
\eea   

\noindent {\bf Schwarzschild-AdS black hole:}\\
 In this case, the metric coefficient and the surface gravity are given by
\bea
g(r) &=& \ls 1-\frac{r_{h}}{r}\rs \lb1-\frac{\L}{3}\ls r^2 + rr_{h} + r^{2}_{h} \rs\rb , \la{eq:metricAdS}\\
\kappa &=& \frac{1}{2r_{h}}(1-\L r^{2}_{h}), \la{eq:surfgravAdS}
\eea 
where $r_{h}$  satisfies
\be
2M = r_{h} \ls 1-\frac{r_{h}^{2}\L}{3} \rs
\ee
and $\L < 0$ is a cosmological constant.
Substituting the above equations into Eqs.(\ref{eq:entropyupto2-scalar})-(\ref{eq:entropyupto2-mx}),  we obtain
\bea
S^{-}_{sc} &=&  \frac{A_{h}}{4\ell^{2}_{p}} - \lb\frac{1}{60}- \frac{\L}{80\pi}  A_{h} \rb
 \ln\ls\frac{A_{h}}{\ell^{2}_{p}}\rs , \la{eq:entropyscAdS}\\
 S^{+}_{wf} &=&  \frac{A_{h}}{4\ell^{2}_{p}} - \lb\frac{23}{96}-  \frac{471}{360} \ls\frac{\L A_{h}}{32\pi}\rs \rb
  \ln\ls\frac{A_{h}}{\ell^{2}_{p}}\rs , \la{eq:entropywfAdS}\\
 S^{-}_{ph} &=&  \frac{A_{h}}{4\ell^{2}_{p}} - \lb\frac{131}{144}- \frac{61}{240}\ls\frac{\L A_{h}}{\pi}\rs \rb
 \ln\ls\frac{A_{h}}{\ell^{2}_{p}}\rs . \la{eq:entropymxAdS}
\eea
Unlike the Schwarzschild case,  the coefficients of logarithmic terms depend on the area.    

\section{Summary}
In this work, we have computed the canonical entropy for a massless quantum field with arbitrary spin
propagating in $3+1$ dimensional spherically symmetric black hole background up to the second order in the 
WKB approximation. The generic structure of the leading and second order expressions for the free energy as 
well as entropy remains the same. However, the second order term contributes significantly to the entropy than the
 leading order. The total entropy up to the second order was obtained by combining the leading and second order 
 results. The total entropy obtained in this manner contains quadratic as well as logarithmic
 divergent parts. These divergences can be cured by renormalizing the gravitational coupling constant 
 \cite{meyers, winstanley}. However, the coefficient for the renormalized gravitational constant is found
 to be different for different species of matter fields \cite{fursaev1}. Consequently, the proper cut-off
 distance of the brick wall from the horizon depends upon the type of field. Thus, by choosing the invariant 
 cut-off appropriately, we were able to express the total entropy as a combination of the standard 
Bekenstein-Hawking  entropy ($S_{BH}$) and the logarithmic correction. The logarithmic contribution to 
the black hole entropy have  been discovered earlier in several different approaches such as \cite{sen3,ashtekar,
partha,carlip1,bibhas1,solodukhin1}.  However, the coefficient of the logarithmic term is generally 
found to be different for different methods. In our case,  the prefactor of the logarithmic term depends on the type of the field. 
For the scalar field,  our result matches with that of  given in  Ref. \cite{sriram1}. It is important to note that these ultra violate 
divergences appearing in our analysis are independent  of the order of the WKB approximation, as  can be easily verified by 
comparing the leading and  the second order expressions. We have also encountered the divergence in the evaluation of 
$\tilde{\l}$ integrals. This divergence occurs near the turning  point of the WKB potential.  However, the structure of these 
additional divergent terms still  prevails irrespective of the additional spin degrees of  freedom of the quantum field. Finally, 
we have applied our analysis to the Schwarzschild and  Schwarzschild-Anti de-Sitter  black holes and obtained the 
expression of the  statistical entropy for the massless scalar, fermion and electromagnetic field, respectively. 
 
  In the present work, we have restricted our computations  up to the second order in the WKB approximation. It
will be interesting to extend the current method and obtain the canonical entropy for quantum fields with the arbitrary 
spins up to $4^{th}$ and $6^{th}$ orders .  Another important aspect that we would like to  investigate in future is the 
thermodynamical stability of the black holes.  In the tunneling mechanism, it was shown  that the inclusion of the quantum 
corrections makes the black holes stable via phase transition Ref.\cite{sujoy2}.  
It will be  worthwhile to study the phase transitions and  thermodynamic stability of the black holes by using the brick wall approach at 
higher orders in the WKB approximation. We would like to address these issues in near future.
\newpage
\noindent {\bf Acknowledgements}\\
We would like to thank E. Son and M. Eune for discussions.  S. Kulkarni  wish to thank  L. Sriramkumar  and S. K. Modak for their useful comments and 
suggestions. S. Kulkarni  is being supported by National Research Foundation (NRF) grant funded by the Korea government (MEST) through the 
Center for  Quantum Spacetime (CQUeST) of Sogang University with grant number 2005-0049409.  W. Kim was  supported by  National 
Research Foundation (NRF) grant funded by the Korea government (MEST) (2012-0002880).

 %%%%%%%%%%%%%%%%%%%%%%%%%%%%%%%%%%%%%%%%%%%%%%%%%%%%%%%
\end{document}